\documentclass[aip,apl,amsmath,amssymb,reprint,superscriptaddress]{revtex4-1}
\usepackage{graphicx}
\usepackage{amsmath}
\usepackage{dcolumn}
\usepackage{bm}
\usepackage{bbold}
\usepackage{tabularx}
\usepackage{xcolor}

\bibstyle{apsrev4-1}

\begin{document}
\title{Generation of Ultra-Low Power Phononic Combs}

\author{Maxim Goryachev}
\email{maxim.goryachev@uwa.edu.au}
\affiliation{ARC Centre of Excellence for Engineered Quantum Systems, University of Western Australia, 35 Stirling Highway, Crawley WA 6009, Australia}

\author{Serge Galliou}
\affiliation{FEMTO-ST Institute, Univ. Bourgogne Franche-Comt\'{e}, CNRS, ENSMM, 26 Rue de l’\'{E}pitaphe 25000 Besan\c{c}on, France}

\author{Michael E. Tobar}
\affiliation{ARC Centre of Excellence for Engineered Quantum Systems, University of Western Australia, 35 Stirling Highway, Crawley WA 6009, Australia}

\begin{abstract}

We demonstrate excitation of phononic frequency combs in a Bulk Acoustic Wave system at a temperature of $20$mK using a single tone low power signal source. The observed ultra low power threshold is due to a combination of very high quality factor of $4.2\times 10^8$ and relatively strong nonlinear effects. The observed repetition rate of the comb varies from 0.7 to 2Hz and spans over tens of Hertz. The demonstrated system is fully excited via piezoelectricity and does not require mode spectra engineering and external optical or microwave signals. It is shown that the comb profile significantly depends on geometry of excitation and detection electrodes. Observed strong Duffing nonlinearity below the generation threshold suggests that the system is a phononic analogue to Kerr frequency combs excited in monolithic optical microresonators. The ultra-low power regime opens a way of integrating this phononic system with quantum hybrid systems such as impurity defects and superconducting qubits.

\end{abstract}

\date{\today}
\maketitle

\section{Introduction}

Photonic frequency comb generators are an important tool used for many scientific applications allowing to couple distant parts of frequency spectrum\cite{K.:2016aa}. Amongst others, these important applications include ultra stable clocks and frequency metrology\cite{Ma:2004aa,Holzwarth:2001aa,Millo:2009aa} and spectroscopy\cite{Picque:2019aa,Stowe:2008aa}. All of these system are photonic mostly utilising a carrier signal at an optical frequency and RF or microwave repetition rates. The most popular ways to generate such combs include mode-locked lasers\cite{Adler:2004aa}, (four)-wave mixing/Kerr nonlinearity\cite{Sefler:1998aa,Creedon2012b, Pasquazi:2018aa,DelHaye:2007ab} or other physical sources of nonlinearities\cite{Maksymov:2019aa}. Despite the fact that phononic systems also play an important role in metrology and spectroscopy, demonstration of phononic or acoustic wave combs has been sporadic\cite{Cao:2014aa}. Moreover, their complexity, requirements to design a certain mode structure, very high threshold powers\cite{Ganesan:2017aa} and requirement for extra laser sources\cite{Hase:2013aa} make these systems inapplicable for many low energy applications. In these applications Bulk Acoustic Wave (BAW) devices found very extensive use, e.g. frequency metrology\cite{Salzenstein:2010aa}, quantum hybrid systems\cite{Chu:2017aa,Kharel:2018aa,Kotler:2017aa} and fundamental physics tests\cite{Goryachev:2014ac,Singh:2017aa,PhysRevX.6.011018,bushev2019}. 
Additionally both bulk and surface acoustic wave devices are utilised for spectroscopy\cite{Lai:1988aa}, detection and sensors\cite{Czanderna:1984aa,Fogel:2016aa,Johannsmann:2008aa}. Recently nonlinear dynamics of mechanical systems including comb generation has also became a subject of theoretical research\cite{theory1}. All these areas could benefit from designing devices with frequency comb functionality. 

\section{Experimental Setup}

To excite frequency combs at low power levels, one needs a system with low losses and strong nonlinear effects. For these reasons, in this work, we employ a phonon trapping\cite{Stevens:1986aa} SC (Stress Compensated) cut\cite{Kusters:2014mn} quartz BAW cavity\cite{ScRep} operating at 20mK. The device is an SC-cut plano-convex shaped plate of 1mm thickness, 23mm diameter and 300mm radius of curvature. Quartz BAW cavities have demonstrated extremely high values of Quality factors at cryogenic temperatures reaching the level of $8\times10^9$\cite{ScRep,Goryachev:2014ac}. At the same time these devices are able to demonstrate a significant levels of nonlinearities. Traditionally nonlinearities in mechanical system are associated with a Duffing nonlinearity that is a consequence of lattice anharmonicity, i.e. a nonlinear term in  Hook's law. This type of nonlinearity has been extensively studied for many types of mechanical systems\cite{nonlin,Tiers3,nosek}. Other sources of nonlinearities come from thermoelectroelastic effects\cite{Tiersten1} and coupling to ensembles of Two Level Systems (TLS), etc. TLS effects are present in many physical systems operating at milli-Kelvin temperatures, such as superconducting qubits, resonators and amplifiers\cite{Lisenfeld:2015aa,Burnett:2014aa,Anton:2013aa}, dielectric resonators\cite{Creedon:2011aa}, mechanical resonators\cite{Goryachev:2019aa}. Particularly in BAW cavities, the presence of TLSs has been demonstrated through a number of effects, namely, nonlinear losses\cite{Goryachev:2014ac}, magnetic field sensitivity (hysteresis and memory behavior)\cite{Goryachev:2019aa} and non-Duffing nonlinearities in devices with high levels of impurities\cite{quartzJAP}, which become dominant at higher frequencies. Thus, to exhibit  only by the Duffing nonlinearity, we limit the experiment to low order overtones. Particularly, the third overtone of the fast shear mode (also known as the B-mode) at the frequency $f_c = 5.506033$MHz is used to demonstrate the effect.

\begin{figure}[hbt!]
     \begin{center}
            \includegraphics[width=0.4\textwidth]{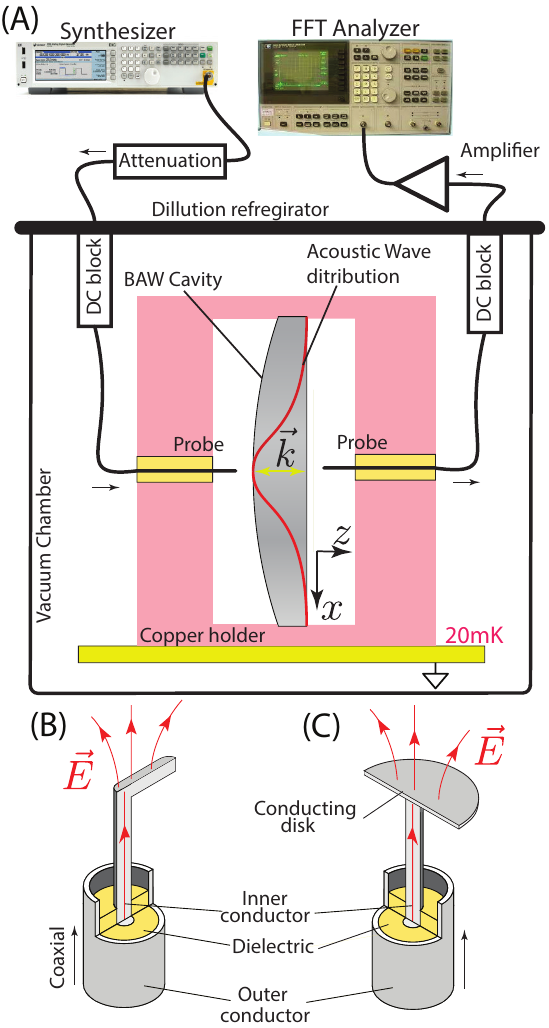}
            \end{center}
    \caption{(A) Experimental setup for excitation of the acoustic frequency comb: a BAW cavity sits vertically in a Copper holder at 20mK in a vacuum chamber. The signal is supplied from the room temperature synthesizer via microwave cables thermalised with DC blocks. The transmitted signal is amplified and analysed with an FFT machine. $\vec{k}$ is the direction of the BAW propagation whose energy has a Gaussian distribution in the crystal plane (red curve). (B) L-shaped excitation antenna (probe) coupling to electric field $\vec{E}$. (C) Excitation antenna with a disk.}%
   \label{setupSIG}
\end{figure}

The experimental setup used in this work is shown in Fig.~\ref{setupSIG} (A). The BAW cavity is placed inside a copper holder attached to a base plate of a dilution refrigerator cooled to 20mK. The internal volume of the holder is 22mm in diameter (with 1mm groove to hold the quartz plate) and 10mm high. The holder is designed to have its lowest frequency microwave resonances above a few GHz to avoid any coupling to any BAW acoustic modes (limited to hundreds of MHz). The holder is not hermetically sealed, thus the device is subject to the vacuum level of the dilution refrigerator, which is about $3\times10^{-6}$mbar. The acoustic modes are excited and read out piezoelectrically using two specially designed coaxial probes coupled to the electric field of the structure. L-shaped and disk-shaped antennas are attached to the central conductor of the co-axial cable protruding from the outer coating by few millimetres and are banded as shown in Fig.~\ref{setupSIG} (B). This is different from the traditional way of excitation of such devices where one uses relatively large flat electrodes creating an effective coupling capacitor. The probes are positioned to point directly at the centre of the BAW crystal. The geometry of the antennas is optimised to increase the transmission signal. The device is excited using a single-tone RF signal synthesiser locked to a Hydrogen maser to achieve the best frequency stability. The power of the incident signal is controlled via a chain of room temperature attenuators. Low frequency components of the excitation signal are removed with a chain of cold DC blocks with a cut off frequency of a few kHz. The transmitted signal is amplified with a low noise amplifier and detected using an FFT (signal) Analyser. 

\section{Comb Observation}

\begin{figure}[hbt!]
     \begin{center}
            \includegraphics[width=0.5\textwidth]{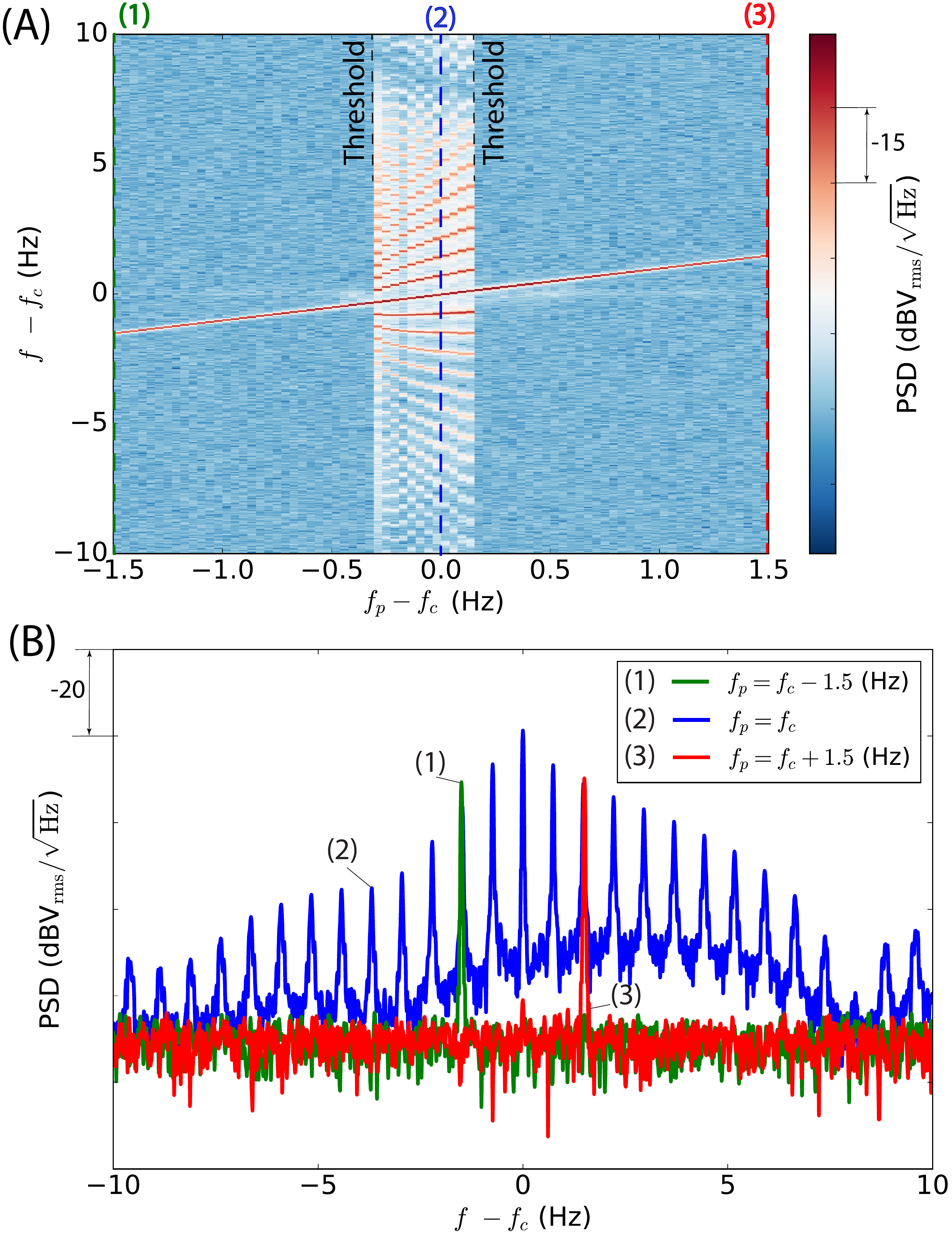}
            \end{center}
    \caption{(A) Output signal PSD $S(f-f_c)$ as a function of pump frequency detuning $f_p - f_c$ for a constant incident power $P=-66$dBm. (B) Three PSDs for different constant incident frequencies $f_p$ and the same power $P$ corresponding to slices of (A) along the $y$ axis.}%
   \label{resp1}
\end{figure}

In the first experiment, the system response in terms of signal Power Spectral Density (PSD) {\color{black}$S(f)$} is measured as a function of the pump frequency $f_p$ in the vicinity of the acoustic resonance $f_s$ for the constant signal power $P$. The result is presented in Fig.~\ref{resp1} where each vertical slice of subplot (A) shows a single PSD trace of the output signal for a constant incident signal frequency $f_p$. Note that each PSD is taken independently for a large number of averages with sufficient time delay between two consecutive measurements. The measurement setup is stable on these time scales because it is locked to an atomic frequency standard. The BAW device parameters are not sensitive to possible temperature fluctuations and other effects. This method ensures the static picture of the response.
Fig.~\ref{resp1} demonstrates a frequency comb when the pump signal frequency approaches that of the resonance. The comb shows two thresholds on each side of the resonance. The repetition rate of the comb is about $0.8$Hz for $f_p=f_c$, slightly increasing with the pump frequency. 

\begin{figure}[hbt!]
     \begin{center}
            \includegraphics[width=0.5\textwidth]{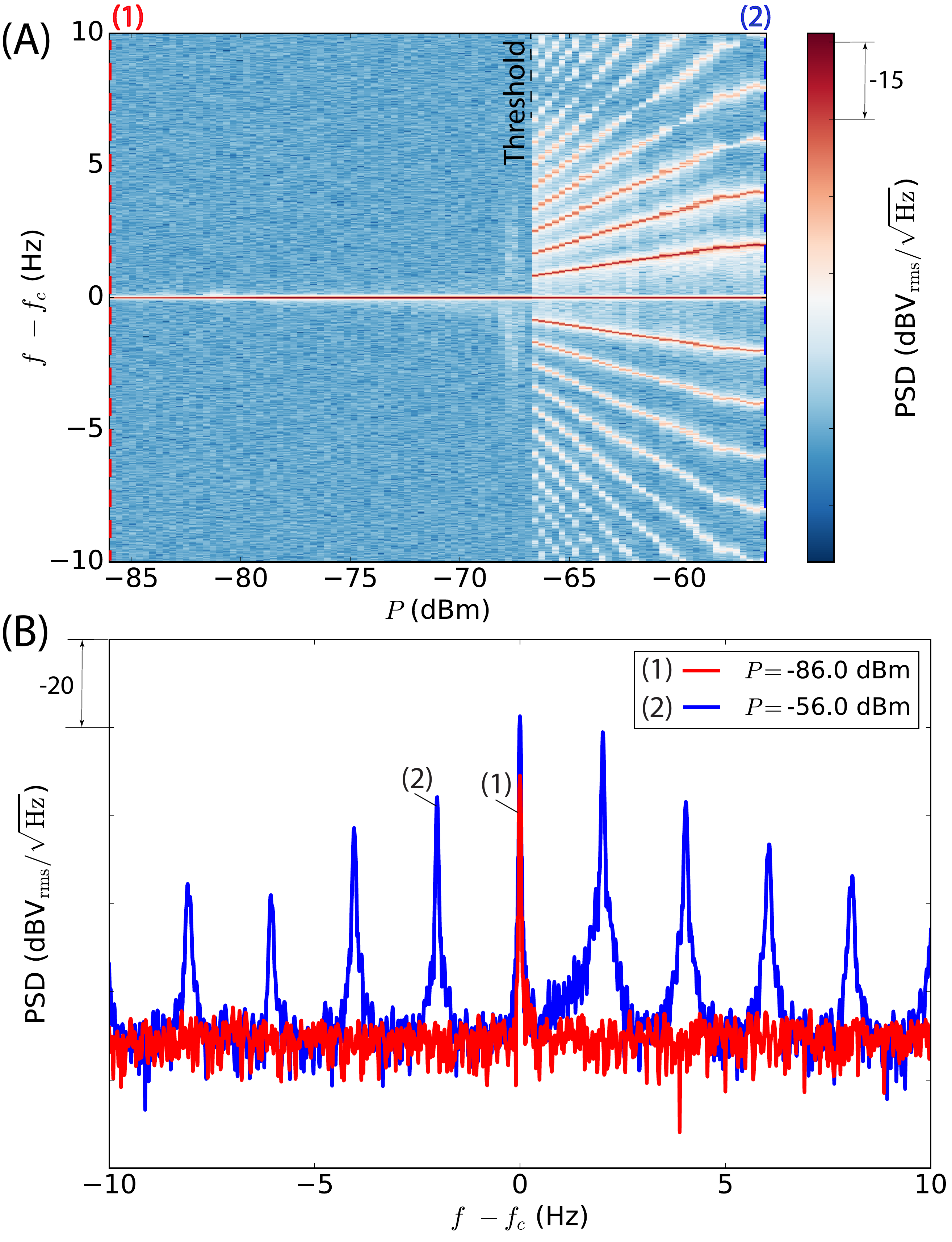}
            \end{center}
    \caption{(A) Output signal PSD $S(f-f_c)$ as a function of pump signal power $P$ for a constant incident frequency $f_p\approx f_c$.  (B) Two PSDs for different constant incident frequency $f_p$ and different power levels $P$ corresponding to slices of (A) along the $y$ axis.}%
   \label{respP1}
\end{figure}

The same threshold in comb generation is observed by varying the incident power for the fixed excitation frequency $f_p\approx f_c$, which is shown in Fig.~\ref{respP1}. The figure also demonstrates a decrease of the repetition rate with power. 

\begin{figure}[hbt!]
     \begin{center}
            \includegraphics[width=0.5\textwidth]{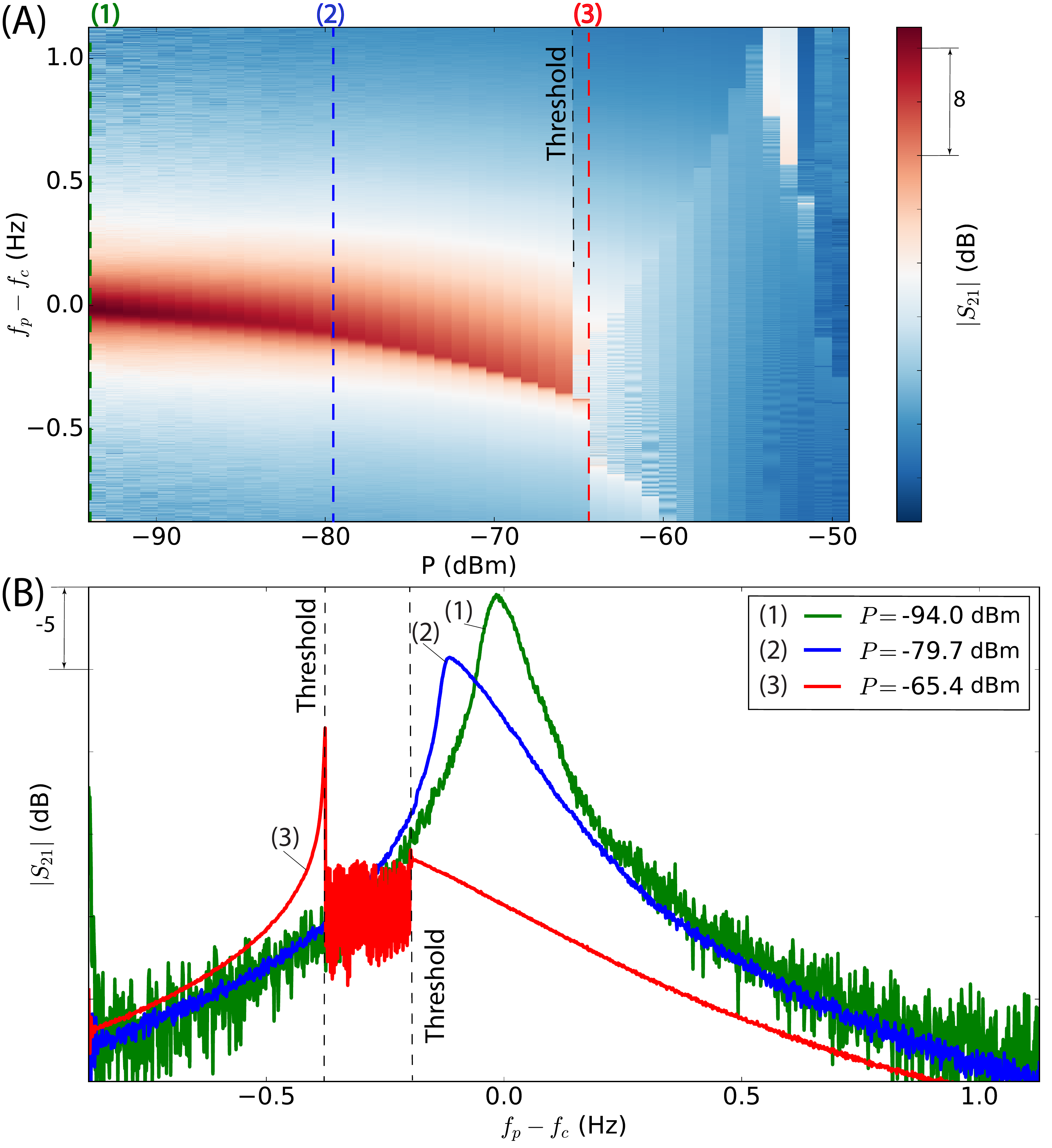}
            \end{center}
    \caption{(A) Transmission through the BAW cavity $|S_{21}(f-f_c)|$ for different values of the incident power $P$. (B) Three examples of transmission for different constant power levels corresponding to slices of (A) along the $y$ axis.}
   \label{transmission}
\end{figure}

In addition to the PSD of the transmitted signal, one may measure transmission directly in terms of scattering parameters $S_{21}$. For this the signal source and the FFT analyzer are replaced with a network analyzer. For this kind of measurement, one sweeps the frequency of the pump signal and measures the ratio of the transmitted and incident wave. To avoid any distortion due to ringing effect, the sweep time for one trace is taken to be 30 minutes. The result is shown in Fig.~\ref{transmission}, where for low incident power a classical Duffing behaviour is observed. After a bifurcation, the system demonstrates irregular behavior that corresponds to the comb generation regime. In this area between two 'Threshold' labels, the output signal has more than one frequency component; thus, the transmission as revealed by a receiving detector becomes modulated resulting in this artefact. From these measurements, in the linear regime, the Quality factor of the resonance is estimated as $4.2\times 10^8$. Fig.~\ref{transmission} also demonstrates the importance of the Duffing nonlinearity below the generation threshold that can be seen even for the lowest excitation power. This observation draws an analogy with the photonic frequency comb based on Kerr nonlinearity (due to the third-order $\chi^{(3)}$ component of the electric susceptibility)\cite{DelHaye:2007ab}, which is a photonic analogue of the phononic Duffing nonlinearity. In both systems, the ultra-low generation threshold is achieved by a combination of nonlinear effects in a low loss environment. 

\section{Comb Simulation}

To understand the comb generation effects in the given acoustic system, one may draw certain analogies and point out some differences with the optical counterparts. Based on the nonlinear elastic equations for a phonon trapping cavity (see Appendix~\ref{uno}), one may deduce an equation of motion for a complex envelope of the displacement signal $U_x(z)$ (displacement along in-plane coordinate $x$ propagating along the thickness coordinate $z$):
\begin{multline}
%\begin{aligned}
\label{GRD112}
	\displaystyle  \partial_{\tau}U_x =-(\kappa + i\Omega) U_x - i\alpha \partial_z^{2}U_x - i\gamma\partial_z\Big[\big|\partial_z U_x\big|^2\partial_z U_x\Big] + F.\\
%\end{aligned}
\end{multline}
Here $\Omega$ is a scaled resonance frequency, $\kappa$ is a loss coefficient, $F$ is a complex amplitude of external excitation, and $\alpha$ and $\gamma$ are frequency scaled stiffness and nonlinear coefficients. The right-hand side of this equation consists of the following terms: resonance frequency (complex and real parts, i.e. damping and central frequency), dispersion term, nonlinear Duffing term and external forcing.
This equation is almost identical to the Lugiato–Lefever equation\cite{Lugiato:1987aa,Castelli:2017aa,K.:2016aa}, which captures dynamics of the Kerr frequency combs in optical systems. The major difference is in the nonlinear terms, Kerr and Duffing, where one is given as a function of the complex envelope and the other is given in terms of its first space derivative. %Due to similarity of equation of motion for acoustic and electromagnetic cavities as well as identical types of nonlinearity, simulation of phononic combs is similar to that of photonic and can be found elsewhere\cite{K.:2016aa}. 

The major difference between optical Kerr combs and the presented acoustic is the absence of a free spectral range corresponding to the comb signal spacing. In principle, absence of the equally spaced mode structure is also a feature of combs in other mechanical systems\cite{Ganesan:2017aa} and in Whispering Gallery Mode crystals\cite{Creedon2012b,Nand:2014aa}. In the latter case, an extra half degree of freedom arises due to the Fano resonance, the phenomenon which in the BAW community is known as anti-resonance\cite{VigQuartz}. This typical additional spectral structure is formed by the electrodes spurious capacitance and acoustic equivalent inductance. %{\color{black} The frequency relation between acoustic resonance and its anti-resonance is described in Appendix~\ref{duo} where it is shown that the frequency ifference between the two resonances is set by a shunt (electrode specific) capacitance. Also, Appendix~\ref{duo} demonstrates generation of a frequency comb in a simple  model a BAW device with a shunt capacitance and Duffing nonlinearity by direct numerical simulation.}

\begin{figure}[hbt!]
     \begin{center}
            \includegraphics[width=0.5\textwidth]{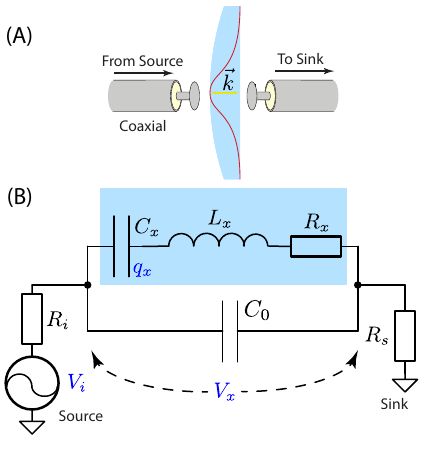}
            \end{center}
    \caption{(A) BAW cavity coupled to input and output transmission lines via disk probes. (B) Equivalent circuit of such device. The blue block represents the pure mechanical degree of freedom. }%
   \label{vanDyke}
\end{figure}

To understand the relationship between two types of resonances in the system, we consider a simplified model of a Bulk Acoustic Cavity with additional parasitic capacitance. Such capacitance arises due to excitation electrodes in all types of BAW devices. For an external signal it is composed of parasitic bypass short-circuiting mechanical resonances. To capture the effect of this capacitance and demonstrate how it leads to two mode dynamics of the system, we employ the well known Butterworth-van Dyke model of a BAW device shown in Fig.~\ref{vanDyke}. In this model, mechanical motion of the system is modelled using motional capacitance $C_x$, inductance $L_z$ and loss resistance $R_x$. The shunt capacitance is represented by $C_0$. The system is coupled to the signal source $V_i$ with loss element $R_i$ and the sink $R_s$. This is the simplest model of the BAW cavity which does not capture all effects but could be used to demonstrate generation of a photonic comb. 

The linear model in Fig.~\ref{vanDyke} (B) is described by the following set of differential equations:
\begin{multline}
\begin{aligned}
	\label{GRN114}
	\displaystyle  L_x\ddot{q}_x + R_x \dot{q}_x + C_x^{-1}q_x = V_x ,\\
	\displaystyle (R_s + R_i) (\dot{q}_0 +  \dot{q}_x) + V_x=V_i,\\
	\displaystyle q_0 = C_0V_x,
\end{aligned}
\end{multline}
where $q_x$ is a charge on equivalent mechanical capacitor, $q_0$ is a charge on shunt capacitor and $V_x$ is a voltage across the device. Equivalently, the system may be described as follows:
\begin{multline}
\begin{aligned}
	\label{GRN115}
	\frac{d}{d\tau}\left( \begin{array}{c} q_x \\ i_x \\ v_x \end{array} \right) = \begin{bmatrix} 0 & 1 & 0 \\ -\omega_0^2 & -\gamma_i & g  \\ 0 & -g & -\gamma_v \end{bmatrix} \left( \begin{array}{c} q_x \\ i_x \\ v_x \end{array} \right) + \left( \begin{array}{c} 0 \\ 0 \\ f \end{array} \right) 
\end{aligned}
\end{multline}
where $\gamma_i = R_x/L_x$,  $\gamma_v = 1/(R_i+R_s)/C_0$, $\omega_0^2 = 1/(C_xL_x)$, $g^2 = 1/(C_0L_x)$, $v_x =  V_x/(gL_x)$ and $v_i=V_i\gamma_z/(L_xg)$.

\begin{figure}[hbt!]
     \begin{center}
            \includegraphics[width=0.5\textwidth]{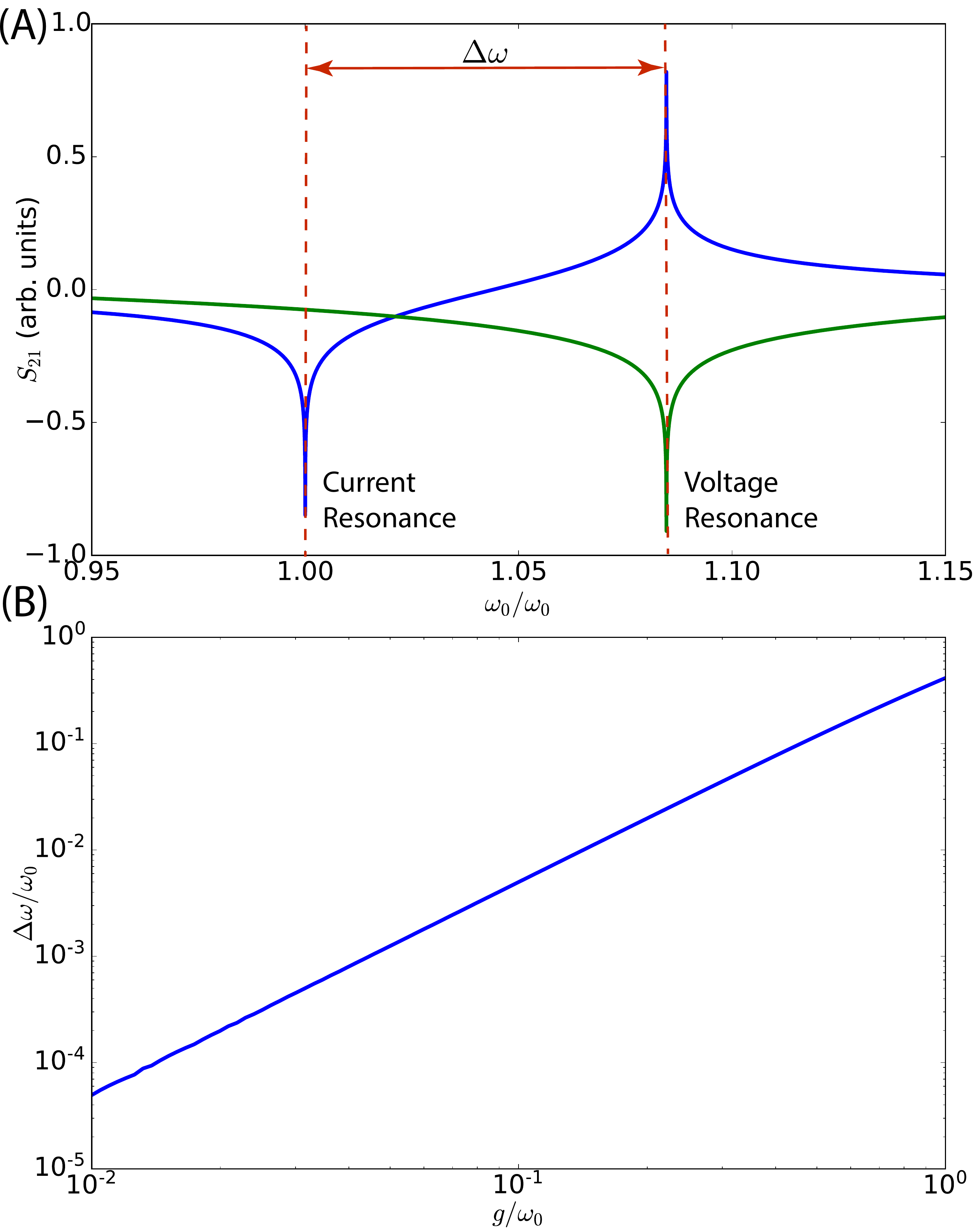}
            \end{center}
    \caption{(A) System response in current and voltage configurations. (B) Scaled frequency difference between current and voltage resonances of $\Delta\omega$ as a function of coupling $g/\omega_0 = \sqrt{C_x/C_0}$. }%
   \label{ZeosPoles}
\end{figure}

Resonance of the system (\ref{GRN115}) can be understood as an equivalent voltage resonance or antiresonance: one imposes voltage $v_i$ over the plate and detect the current $i_x$. It corresponds to poles of the $3\times3$ matrix on the left-hand side of this system. The other type of resonance is an equivalent current corresponding to the situation when one injects current and detects voltage. This situation corresponds to zeros of the $3\times3$ matrix on the left hand side. The frequency difference between the two increases with increasing shunt capacitance. This is demonstrated in Fig.~\ref{ZeosPoles} where the frequency difference between zeros and poles of the system is shown as a function of parameter $g/\omega_0 = \sqrt{C_x/C_0}$. In fact, this parameter can be very small due to smallness of equivalent motional capacitance $C_x$. This can render the frequency difference to a few Hz. Moreover, the system can be understood as a strongly coupled “resonance-antiresonance” system where equivalent voltages and currents are mixed through the strong Duffing nonlinearity. This model also explains change of the comb spacing with the applied power by the fact that the resonance frequency shifts with respect to the anti-resonance according to the Duffing model.

\begin{figure}[hbt!]
     \begin{center}
            \includegraphics[width=0.5\textwidth]{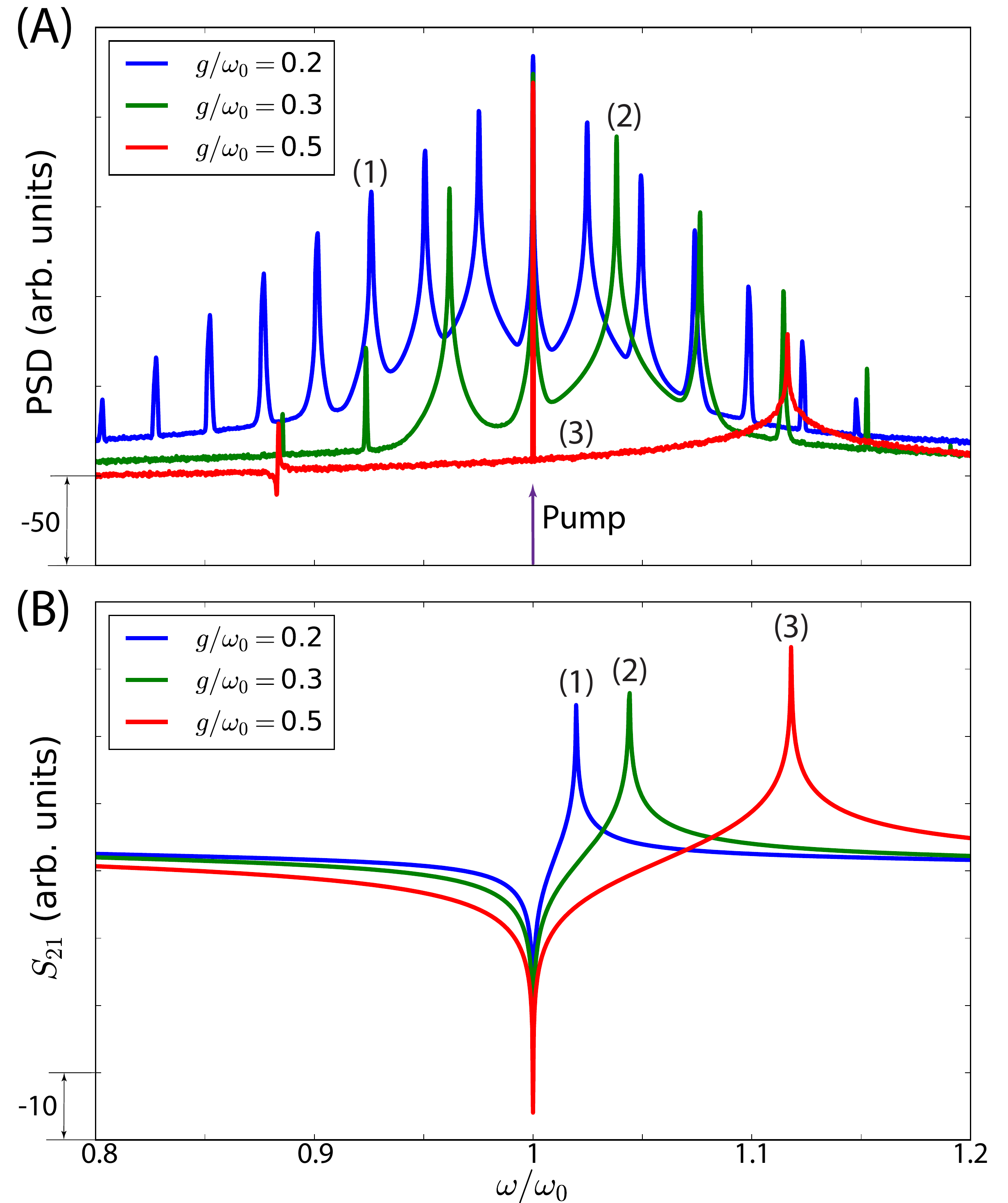}
            \end{center}
    \caption{(A) Power Spectral Density of a steady state signal $i_x$ obtained by numerical simulation of the BAW model with shunt capacitance $g/\omega_0$ and Duffing nonlinearity. (B) Corresponding transmission responses. }%
   \label{simulation}
\end{figure}

To demonstrate the ability of the system to generate a comb, we add a Duffing term $\kappa q_x^3$ into the second equation of motion in system (\ref{GRN115}) and simulate the resulting system of nonlinear equations. The simulation is done in the time domain, Power Spectral Densities (PSDs) are calculated for steady state regimes. Stability of the solutions is verified by changing time steps and tolerances of the simulations. The two loss rates are chosen to be much less than $g$, i.e. $\gamma_i = 10^{-6}\omega_0$ and $\gamma_i = 10^{-4}\omega_0$. An external signal is pumped at frequency $\omega_0$. Note should be taken that all simulation parameters are chosen for demonstration purposes and only roughly correspond to the parameters of the actual system. The resulting PSDs for different values of scaled coupling $g/\omega_0$ are presented in Fig.~\ref{simulation} (A). In this Figure, the comb repetition rate is clearly linked to the spacing between resonance and antiresonance in Fig.~\ref{simulation} (B), and affirms the experimental results presented in this work.

\section{Effect of Electrodes}

It has to be emphasised that the comb generation cannot be related to spurious mode coupling. The mode spectrum of the quartz BAW resonators of this kind is well known both from theoretical calculations\cite{Stevens:1986aa} and from experiment, e.g. X-ray topography\cite{Irzhak:2007aa}. The closest modes of the same family with a different in-plane mode numbers are situated at least several kHz away, which comprises a few tens of thousands linewidths. The shear mode with orthogonal polarisation is below 5~MHz. Also, similar comb effects are observed for other shear modes as well. Although the observed generation threshold is higher due to lower Quality factors of the corresponding modes. The longitudinal modes exhibit higher resonance frequencies due to a significantly different wave velocity lacking the comb generation. This may be attributed to the fact that these modes exhibit a very different type of nonlinearity\cite{Goryachev:2014ac,Goryachev:2019aa}.  

\begin{figure}[hbt!]
     \begin{center}
            \includegraphics[width=0.5\textwidth]{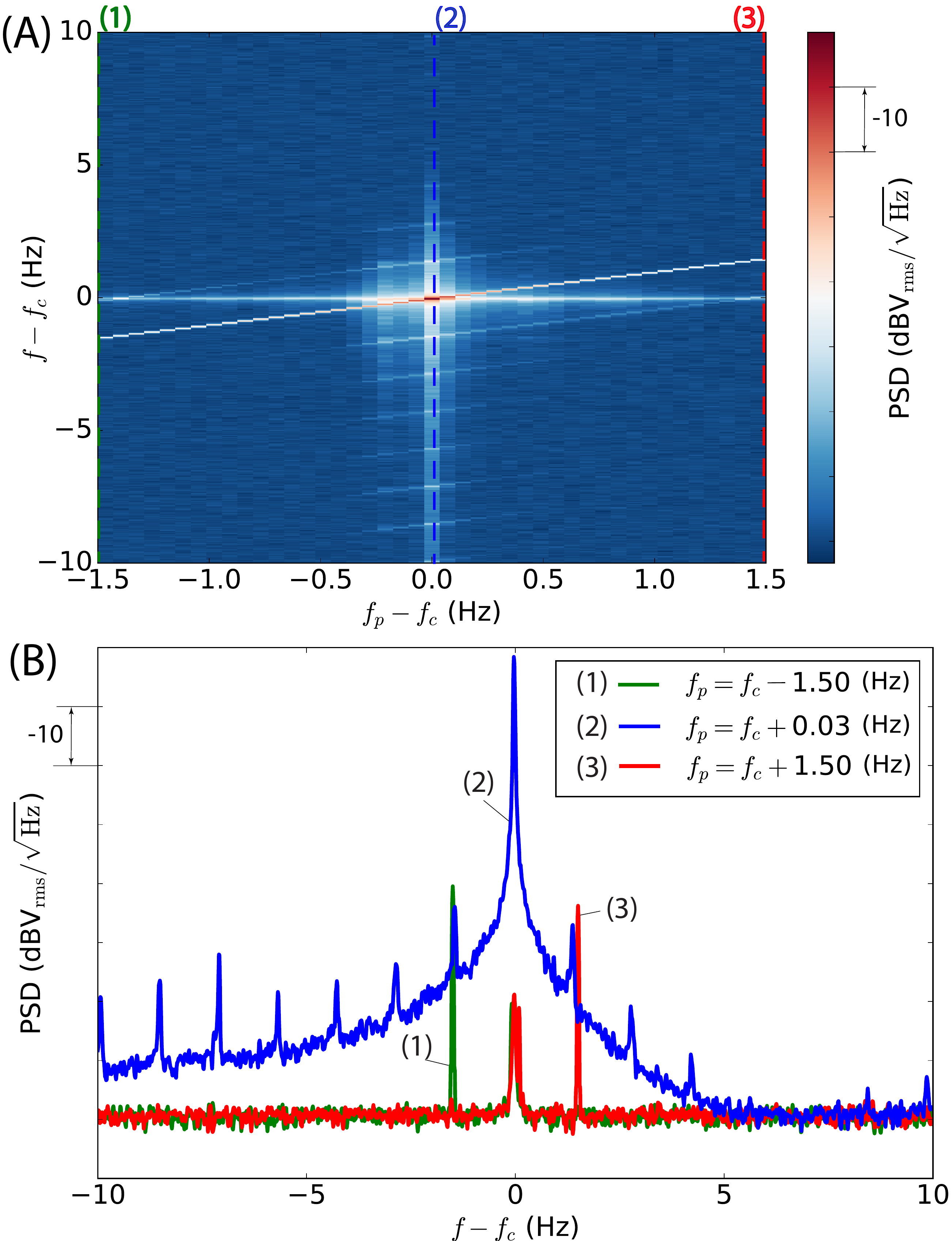}
            \end{center}
    \caption{(A) Output signal PSD $S(f-f_c)$ as a function of pump frequency detuning $f_p - f_c$ for a constant incident power $P=-80$dBm for a system excited with a metallic disk antennas. (B) Three PSDs for different constant incident frequency $f_p$ and the same power $P$ corresponding to slices of (A) along $y$ axis.}
   \label{resp2}
\end{figure}

As suggested above, the observed comb generation is related to the anti-resonance phenomenon, and thus, on the system response should significantly depends on the geometry of the excitation electrodes. To verify this statement the system has been excited with a different type of antenna. In the previous results, the system was excited using L-shaped asymmetric probes. A qualitatively different result is observed when both probe antennas have a metallic disk (1cm diameter sticking 4mm from the copper surface) at their end which is shown in Fig.~\ref{setupSIG} (B). The system response for such configuration is shown in Fig.~\ref{resp2}. Here, the comb appear to be at lower incident power level ($-80$dBm) with clear one-sided character. The comb repetition rate is about $1.5$~Hz and can be adjusted in the range between $0.7$ and $2$~Hz. Due to very low level of the output signal, these results do not display a sharp threshold as it occurs below the noise floor. Another interesting phenomenon shown in this figure is the increase of noise in the generation regime (curve (2)). Such excess noise in BAW devices and its dependence on power levels have been a subject of investigation for many decades as it limits the performance of these devices at room temperature\cite{Gagnepain:1983aa,Boudot:2016aa}. Although the origins of such phenomena are still unknown.

\section{Conclusion}

In summary, we experimentally demonstrated excitation of ultra-low threshold phononic frequency combs in a cryogenic environment. The result is achieved at very low acoustic loss and strong nonlinearities. The comb generation is supported by the anti-resonance phenomenon due to electrodes that is demonstrated by numerical simulation. The comb profile could be engineered with the electrode geometry. On the other hand the method does not require special designs of mode spectra or external microwave or optical signals. The system demonstrates similarities with optical Kerr effect frequency comb generators based on optical microresonators. Because of its ultra low power, the system becomes compatible with quantum hybrid systems\cite{Chu:2017aa,Kharel:2018aa,Kotler:2017aa}, particularly superconducting and impurity defects qubits, and can be used for their manipulation\cite{Hann2019, Chen:2019aa}. Other applications include sensors, phonon based computations\cite{Kazmi:2017aa} and tests of fundamental physics tests\cite{Goryachev:2014ac,Singh:2017aa,PhysRevX.6.011018,bushev2019}. Finally, such systems may be used to improve the BAW device technology through deeper understanding of fundamental processes in solid state acoustics and limiting mechanisms such as the drive level sensitivity and excess noise\cite{Gagnepain:1983aa,Boudot:2016aa}.

This work was supported by the Australian Research Council Grant No. CE170100009.

\section*{References}
%\bibliography{biblioBAW}

%merlin.mbs aipnum4-1.bst 2010-07-25 4.21a (PWD, AO, DPC) hacked
%Control: key (0)
%Control: author (8) initials jnrlst
%Control: editor formatted (1) identically to author
%Control: production of article title (-1) disabled
%Control: page (0) single
%Control: year (1) truncated
%Control: production of eprint (0) enabled
%

\appendix 

\section{Elastic Nonlinearity in a BAW Cavity}
\label{uno}

To discuss nonlinearities in a curved (phonon trapping) BAW cavity, we consider a crystal plate oriented in such a way that the the crystal thickness is along coordinate $z$ (see Fig.~\ref{setupSIG}). Nonlinear Equations of Motion (EOM) for mechanical displacement $\mathbf{u} = (u_x, u_y, u_z)^T$ of a shear mode with $u_x$ being the major component ($|\partial_z u_i|\gg|\partial_x u_i|,|\partial_y u_i|$) are presented as follows\cite{Tiers2}: 
\begin{multline}
\begin{aligned}
	\label{GRN111}
	\displaystyle  \rho \partial_t^{2}{u}_x - \overline{c}^{(1)} \partial_z^{2}u_x - \widetilde{\gamma}\partial_z\big(\partial_z u_x\big)^3 - \widetilde{\beta}_1 \partial_z\big(\partial_z u_x\big)^2 \\
	\displaystyle - 2\widetilde{\beta}_2 \partial_z\big(\partial_z u_x\partial_z u_z\big) - 2\widetilde{\beta}_2  \partial_z\big(\partial_z u_x\partial_z u_y\big)= f_x,\\
	\displaystyle  \rho \partial_t^{2}{u}_y  - \overline{c}^{(2)} \partial_z^{2}u_y - \widetilde{\beta}_2  \partial_z\big(\partial_z u_x\big)^2 = f_y,\\
	\displaystyle  \rho \partial_t^{2}{u}_z  - \overline{c}^{(3)} \partial_z^{2}u_z - \widetilde{\beta}_2  \partial_z\big(\partial_z u_x\big)^2 = f_z,
\end{aligned}
\end{multline}
where $\rho$ is the specific mass, $\overline{c}^{(i)}$ are so-called piezoelectrically stiffened eigenvalues, $f_i$ are forcing signals, $\widetilde{\gamma}$ and $\widetilde{\beta}_i$ are nonlinear coefficients. For these equations, one neglects space derivatives with respect to in plane coordinates $x$ and $y$ as variation along these axes is significantly smaller than along the thickness axis $z$. 

Equations (\ref{GRN111}) include two types of nonlinearities: self nonlinearities (third and second order in terms of EOMs) of the main component of displacement $u_x$, cross nonlinearities between displacements in different directions (all second order). It is widely accepted that the third order nonlinearities in terms of EOMs are the dominant for solids\cite{Tiers4,Gagnepain:1983aa,nosek}, thus for the first simplest model all $\beta$ coefficients can be neglected. This leaves only one EOM for the major component $u_x(z)$. 

By going into the space of slowly varying (both in space and time) complex amplitude $U_x(\tau,z)$, one can transform these equations into the complex form:
\begin{multline}
%\begin{aligned}
\label{GRN112}
	\displaystyle  \partial_{\tau}U_x =-(\kappa + i\Omega) U_x - i\alpha \partial_z^{2}U_x - i\gamma\partial_z\Big[\big|\partial_z U_x\big|^2\partial_z U_x\Big] + F,\\
%\end{aligned}
\end{multline}
where $\Omega$ is a scaled resonance frequency, $\kappa$ is the loss coefficient, $F$ is a complex amplitude of external excitation, $\alpha$ and $\gamma$ are frequency scaled stiffness and nonlinear coefficients. The right-hand side of this equation consists of following terms: resonance frequency (complex and real parts, i.e. damping and central frequencies), dispersion term, nonlinear Duffing term and external forcing.

Equation (\ref{GRN111}) describing spatial and temporal evolution of nonlinear acoustic excitation is analogous to the Lugiato–Lefever equation\cite{Lugiato:1987aa} used to describe Kerr frequency combs in optical microresonators\cite{Castelli:2017aa,K.:2016aa}:
\begin{multline}
%\begin{aligned}
	\label{GRN113}
	\displaystyle  \partial_{\tau}E =-(\kappa + i\Omega) E - i\alpha \partial_z^{2}E - i\gamma\big| E\big|^2 E + F,\\
%\end{aligned}
\end{multline}
where $E$ is the complex amplitude of the electrical field. Similar to acoustic equation (\ref{GRN112}), the right-hand side of the Lugiato–Lefever equation consists of resonance frequency (complex and real parts, i.e. damping and central frequencies), dispersion, nonlinear Kerr and external forcing terms.

The major difference between the photonic and phononic nonlinear equations (\ref{GRN113}) and (\ref{GRN112}) is the nonlinear term where one is written in terms of the complex amplitude $E$ and the other is written in terms of the first spatial derivative of the complex amplitude $U$. The latter arises due to the fact that the nonlinearity originates from the strain rather than displacement itself.  

% {\color{black}\section{Dual Resonance in a BAW Cavity}
%\label{duo}}

\end{document}